\documentclass[twocolumn,amsmath,amssymb,prl,superscriptaddress]{revtex4}
\usepackage{graphicx}
\usepackage{braket} % brakets

\newcommand{\XX}{X\!X}

\begin{document}

\title{On-demand source of maximally entangled photon-pairs using the biexciton-exciton radiative cascade}

\author{R. Winik}
\affiliation{Andrew and Erna Viterbi Department of Electrical Engineering, Technion, Haifa 32000, Israel}
\affiliation{The Physics Department and the Solid State Institute, Technion--Israel Institute of Technology, 32000 Haifa, Israel}
\author{D. Cogan} 
\affiliation{The Physics Department and the Solid State Institute, Technion--Israel Institute of Technology, 32000 Haifa, Israel}
\author{Y. Don}
\affiliation{The Physics Department and the Solid State Institute, Technion--Israel Institute of Technology, 32000 Haifa, Israel}
\author{I. Schwartz}
\affiliation{The Physics Department and the Solid State Institute, Technion--Israel Institute of Technology, 32000 Haifa, Israel}
\author{L. Gantz}
\affiliation{The Physics Department and the Solid State Institute, Technion--Israel Institute of Technology, 32000 Haifa, Israel}
\author{E. R. Schmidgall}
\affiliation{The Physics Department and the Solid State Institute, Technion--Israel Institute of Technology, 32000 Haifa, Israel}
\author{N. Livneh}
\affiliation{Applied Physics Department, The Benin School of computer sciences and engineering, The Hebrew University, Jerusalem 91904, Israel}
\author{R. Rapaport}
\affiliation{Applied Physics Department, The Benin School of computer sciences and engineering, The Hebrew University, Jerusalem 91904, Israel}
\author{E. Buks}
\affiliation{Andrew and Erna Viterbi Department of Electrical Engineering, Technion, Haifa 32000, Israel}
\author{D. Gershoni}
\affiliation{The Physics Department and the Solid State Institute, Technion--Israel Institute of Technology, 32000 Haifa, Israel}
\email[]{dg@physics.technion.ac.il}
\date{\today}

\begin{abstract}
We perform full time resolved tomographic measurements of the polarization state of pairs of photons emitted during the 
radiative cascade of the confined biexciton in a semiconductor quantum dot.  
The biexciton was deterministically initiated using a $\pi$-area pulse into the biexciton two-photon absorption resonance. 
Our measurements demonstrate that the polarization states of the emitted photon pair are maximally entangled. We show that the measured degree
of entanglement depends solely on the temporal resolution by which the time difference between the emissions of the photon pair is determined. 
A route for fabricating an on demand source of maximally polarization entangled photon pairs is thereby provided. 

\end{abstract}

%\keywords{}

\maketitle

The ability to generate entangled photons on-demand is crucial for many future applications in quantum information processing. Devices based on the biexciton-exciton radiative cascade in single semiconductor quantum dot are considered to be one of the best candidates for these applications \cite{benson2000,akopian2006,sellinart2010}. 
The ability to deterministically excite the biexciton using its two-photon absorption resonance \cite{bounouar2015phonon,muller2014} makes this avenue even more promising. A remaining challenge, however, is the excitonic fine structure, which splits the two exciton eigenstates thus providing spectral ``which-path" information on the radiative cascade and preventing the pairs of emitted photons from being polarization entangled \cite{akopian2006}. 
Various strategies were tried in an attempt to reduce the influence of the fine structure splitting. Spectral \cite{akopian2006} and temporal filtering \cite{akopian2006,stevenson2008evolution}, which introduce non desired, non-deterministic post selection. Enhancement of the radiative rate using the Purcell effect \cite{sellinart2010}, thereby reducing, but not limiting the effect of exciton precession. Attempts to reduce the fine-structure splitting using heat treatment \cite{young2005inversion} or growth along the [111] crystalographic direction \cite{versteegh2014observation} were reported as well as applications of external stress \cite{zhang2015high},  electric \cite{bennett2010electric} and magnetic fields tuning \cite{young2006improved,stevenson2006magnetic}. These efforts, usually result in unwanted loss of emission quantum efficiency \cite{bennett2010electric}, and increase in the exciton spin decoherence \cite{stevenson2008evolution}.   

We present here a novel study of a single semiconductor quantum dot, optically depleted~\cite{Schmidgall2015} and then resonantly excited on-demand by a $\pi $-area pulse to the biexciton two photon absorption resonance ~\cite{muller2014}.
The resulting pairs of biexciton and exciton photons are detected by two superconducting detectors synchronized to the exciting laser pulse. By performing synchronized time resolved polarization tomography of the two emitted photons, we unambigously show that the photons remain maximally polarization entangled during the whole radiative decay, and that the measured degree of entanglement does not depend on the QD source, but rather depends on the temporal resolution by which the time difference between the two photon emissions can be determined. Since during the radiative decay the exciton does not lose coherence, there is no need to eliminate the excitonic fine structure splitting. A relatively simple arrangement ~\cite{ jones2006photon, wang2010demand} should therefore provide a reliable source of on demand pairs of maximally entangled photons from a single quantum dot regardless of its non-vanishing excitonic fine structure splitting. 

\begin{figure}[t]
\begin{center}
\includegraphics[width=\columnwidth]{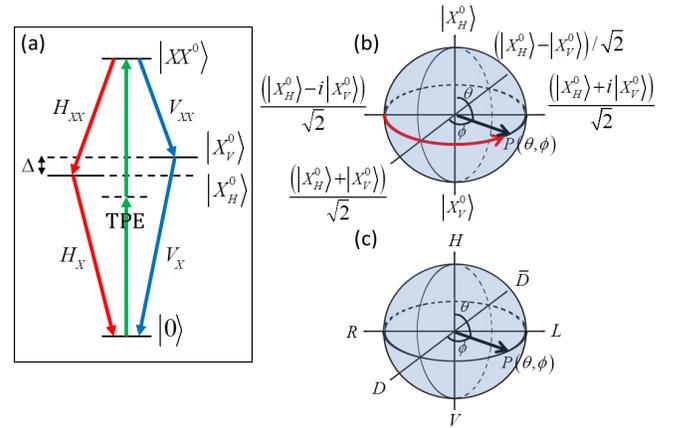}
\caption{\label{fig:spectrum} 
%Two-photon excitation scheme of the biexciton state. 
(a) Schematic description of the energy levels and optical transitions involved in the biexciton two photon resonant excitation (TPE) and its radiative decay cascade. 
(b) The exciton Bloch sphere. The red arrow describes precession of a coherent superposition of its two eigenstates the $X^{0}_H$ and the $X^{0}_V$.
(c) The photon polarization Poincar\'e sphere.
 }
\end{center}
\end{figure}

\section*{Theoretical background}
\begin{figure}[t]
\includegraphics[width=\columnwidth]{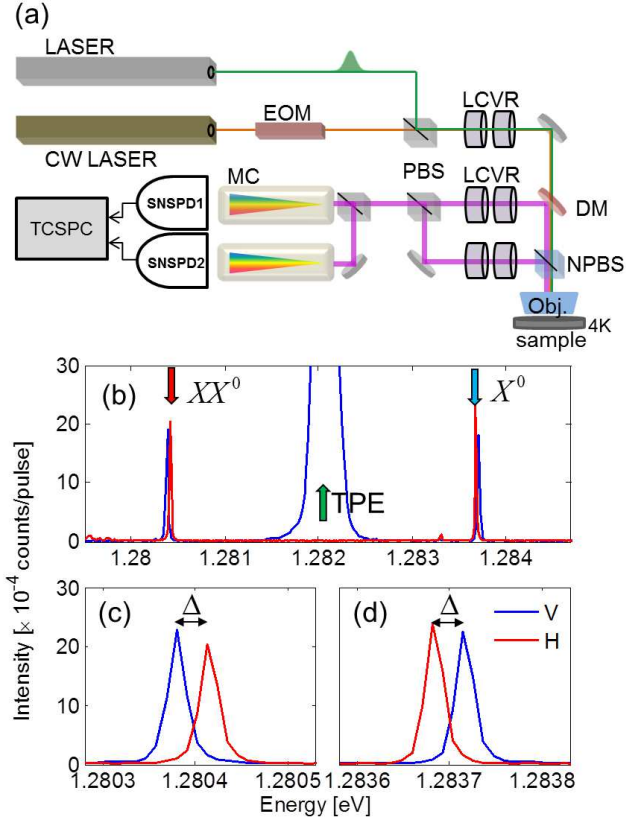}
\caption{\label{fig:Fig2} 
(a) Schematic description of the experimental system.
(b) Polarization sensitive PL spectra of the biexciton-exciton radiative cascade under pulsed excitation into the two photon absorption resonance of the biexciton. Note that for this particular measurement the laser was rectilinearly vertically polarized.  (c), [(d)] expanded energy scale spectra for the biexciton ($XX^0$) [exciton ($X^0$)] spectral lines. (Abbreviations: PBS - polarizing beam splitter, MC - monochromator, SNSPD - superconducting nanowire single photon detector, LCVR - liquid crystal variable retarder, TCSPC - time-correlated single photon counter, EOM - electro optic modulator, DM - dichroic mirror)
} 
\end{figure}
The biexciton-exciton radiative cascade is schematically described in 
Fig.~\ref{fig:spectrum}(a). 
In the figure, resonant excitation of the biexciton using a $\pi$-area pulse to the virtual two photon excitation (TPE) resonance ~\cite{brunner1994sharp, stufler2006two,jayakumar2013deterministic,muller2014}, is described by an upward green arrow. This 12 ps laser pulse follows a few nanosecond long optical depletion pulse~\cite{Schmidgall2015} which empties the QD from charges and long lived dark excitons. The $\pi$-area pulse therefore deterministically photogenerates a confined ground state biexciton ($\ket{XX^0}$) in the QD~\cite{brunner1994sharp,muller2014}. The biexciton spontaneously radiatively decays, leaving in the QD an exciton in a coherent superposition of its two eigenstates: the $\ket{X^{0}_{H}}$ and the $\ket{X^{0}_{V}}$  
states~\cite{gammon1996, kulakovskii1999fine}. The optical selection rules for the biexciton radiative recombination and the lack of information by ``which path" the recombination proceeds result in entanglement between the exciton state and the polarization state of the emitted photon~\cite{benny2011prl,poem2011,ward2014coherent}. 
Their mutual wavefunction is given by:
\begin{equation}
\ket{\psi_{P_1X^0}}=\frac{1}{\sqrt{2}}\left(\ket{H_1X^{0}_{H}}+\ket{V_1X^{0}_{V}}\right)
\end{equation} 
where $\ket{H_1}$ and $\ket{V_1}$ are the two rectilinear polarization states of the first (biexciton) photon.
%The exciton remains in the QD in a coherent superposition of its two eigenstates. 
Since the two exciton eigenstates are not degenerate, the relative phase between these eigenstates precesses in time with a period 
of $T_P=h/\Delta$ where $h$ is the Planck constant and $\Delta$ is the exciton fine structure splitting (see Fig.~\ref{fig:spectrum}(a)). This precession is schematically described on the exciton Bloch sphere in Fig.~\ref{fig:spectrum}(b).
The precession ``stops'' when the exciton recombines and the radiative cascade is completed with the emission of a second photon. 
The two photons are thus entangled. Their mutual wavefunction depends on the recombination time and is given by:     
\begin{equation}
\ket{\psi_{P_1P_2}(t)}=\frac{1}{\sqrt{2}}\left(\ket{H_1H_2}+e^{-i2\pi \frac{t}{T_P}}\ket{V_1V_2}\right)
\end{equation}
where $\ket{H_2}$ and $\ket{V_2}$ are the second (exciton) photon polarization states and $t=t_{X^0}-t_{XX^0}$ is the time between the emission of the biexciton photon $t_{XX^0}$ to that of the exciton $t_{X^0}$.  
It follows that the normalized two photon polarization density matrix is expressed in the basis $\ket{H_1H_2}$, $\ket{H_1V_2}$, $\ket{V_1H_2}$ and $\ket{V_1V_2}$ as: 
\begin{equation}
	\rho_{P_1P_2}(t)=
	%\frac{e^{-\frac{t}{\tau_R}}}{2\tau_R}\begin{pmatrix}
\frac{1}{2}\begin{pmatrix}
		1 & 0 & 0 & e^{-i2\pi \frac{t}{T_P}}\\
		0 & 0 & 0 & 0\\
		0 & 0 & 0 & 0\\
		e^{i2\pi \frac{t}{T_P}} & 0 & 0& 1
	\end{pmatrix}
	\label{eq:eq2}
\end{equation}

To evaluate the degree of entanglement between the two photon polarization states we used a standard measure: the negativity~$\mathcal{N}$ of their polarization density matrix~\cite{peres1996separability}, where~$\mathcal{N}$ is defined as the magnitude of the negative eigenvalue of the partially transposed density matrix. For $\mathcal{N}>0$, the two photons are entangled with fidelity of about $2\mathcal{N}$ to a two photon Bell-state. $\mathcal{N}=\frac{1}{2}$ corresponds to maximal entanglement with unit fidelity to two photon Bell state.
It is easy to see from Eq.  (\ref{eq:eq2}) that the negativity of the two photon density matrix is given by  
\begin{equation}
	\mathcal{N}[\rho_{P_1P_2}(t)]=	
\frac{1}{2}|e^{i\frac{\Delta}{\hbar}t}|=
\frac{1}{2}|e^{i2\pi \frac{t}{T_P}}|
=\frac{1}{2}	
\label{negativity}
\end{equation}
and thus the two photon polarization states are always maximally entangled. 
The information contained in the different colors of the photon pairs, which reveals the decay path of the radiative cascade, remains in the temporal dependence of the phase of the non-diagonal elements of the polarization density matrix.
 
In an actual experiment, the temporal resolution is limited. Thus, if one cannot define the time difference between the two photons to a better resolution than $\Delta T$ the negativity in the measured polarization density matrix is given by:
\begin{equation}
%\begin{eqnarray}
\mathcal{N}[\rho_{P_1P_2}(t,\Delta T)]=
\frac{1}{2\Delta T}\left|\int_{t}^{t+\Delta T}{e^{i2\pi \frac{t'}{T_P}}dt'}\right| 
%\end{eqnarray}
\end{equation}
where for simplicity we assumed that the temporal resolution is represented by a square temporal window of width $\Delta T$. 
It follows that the measured negativity is independent of the time $t$:
\begin{equation}
\mathcal{N}[\rho_{P_1P_2}(\Delta T)]=\frac{1}{2}\left|\text{sinc}\left(\pi\frac{\Delta T}{T_P}\right)\right| 
\label{window}
\end{equation}
%\end{eqnarray}
As can be seen from Eq. \ref{window} the negativity and hence the measured degree of entanglement vanish for $\Delta T=T_P$, but it revives before vanishing for a wider temporal window  ($\Delta T \gg T_P$ ). 

We define a general polarization state of light as a point on the Poincar\'e sphere in which the horizontal, $\ket{H}$ (vertical, $\ket{V}$) rectilinear polarization forms the north (south) pole of the unit sphere as shown in Fig.~\ref{fig:spectrum}(c)~\cite{scully1997quantum}. Thus any polarization can be described in terms of the rectilinear basis using two angles $\theta$ and $\phi$  
\begin{equation}
%\ket{P(\theta,\phi)}=\cos(\theta/2)\ket{H}+e^{i\phi}\sin(\theta/2)\ket{V}
\ket{P(\theta,\phi)}=\cos(\theta/2)\ket{H}+e^{i\phi}\sin(\theta/2)\ket{V}
%P(\theta, \phi)=cos^2(\theta)\left|\ket{H}+e^{i\phi}sin^2(\theta)\left|\ket{V}
\end{equation}
In this representation, the rectilinear polarization basis is given by:
\begin{eqnarray}
\ket{H}=\ket{P(\theta=0,\phi)} \nonumber \\
\ket{{V}}=\ket{P(\theta=\pi,\phi}
\label{EQ6_1}
\end{eqnarray}
and two additional orthogonal bases, the diagonal linear and the circular polarization bases are given by:
\begin{eqnarray}
\ket{D}=(\ket{H}+\ket{V})/\sqrt{2}=\ket{P(\theta=\pi/2,\phi=0)} \nonumber \\
\ket{\bar{D}}=(\ket{H}-\ket{V})/\sqrt{2}=\ket{P(\theta=\pi/2,\phi=\pi)}
\label{EQ6_2}
\end{eqnarray}
and:
\begin{eqnarray}
\ket{L}=(\ket{H}+i\ket{V})/\sqrt{2}=\ket{P(\theta=\pi/2,\phi=\pi/2)} \nonumber \\
\ket{R}=(\ket{H}-i\ket{V})/\sqrt{2}=\ket{P(\theta=\pi/2,\phi=3\pi/2)}
\label{EQ6_3}
\end{eqnarray}
These bases are schematically described in Fig.~\ref{fig:spectrum}(c).

Once the first photon is detected, then the probability rate to detect the first photon with polarization $P_1$ and the second cascading photon in polarization $P_2$ at time $t$ later is given by:
\begin{eqnarray}
p[t,P_1(\theta_1,\phi_1),P_2(\theta_2,\phi_2)]= p_{X^0}(t) \cdot |\braket{P_1P_2|\psi_{P_1P_2}(t)}|^{2}\nonumber\\ 
=\frac{e^{-\frac{t}{\tau_R}}}{2\tau_R}\Bigl|\cos\frac{\theta_1-\theta_2}{2}\cos\left(\frac{\phi_1+\phi_2}{2}+\frac{\pi t}{T_P}\right)+ \nonumber \\
+i\cos\frac{\theta_1+\theta_2}{2}\sin\left(\frac{\phi_1+\phi_2}{2}+ \frac{\pi t}{T_P}\right) \Bigl|^2   \nonumber \\
%\end{pmatrix}\end{equation}
\label{EQ7}
\end{eqnarray}
where $p_{X^0}(t)=e^{-\frac{t}{\tau_R}}/\tau_R$ is the recombination rate of the exciton and $\tau_R$ is the exciton radiative lifetime. 
Note that the probability $P_{HH}$ ($P_{VV}$) that the QD emits both photons co-linearly polarized H (V) is given by 
\begin{eqnarray}
P_{HH(VV)}=\int_0^{\infty} p[t',HH(VV)]dt'=\frac{1}{2}
\end{eqnarray}
as expected for an on-demand generated biexciton radiative cascade. The probability to detect the two photons in a given polarization state $P^{D}_{P_1P_2}$ is given by $P^{D}_{P_1P_2}=\eta^2 P_{P_1P_2}$ where $\eta$ is the light harvesting efficiency of our experimental system~\cite{schmidgall2014deterministic}.

For polarization tomography measurements one detects two photons, which temporally belong to the same cascade, and projects the polarization state of each photon on the three orthogonal polarization bases. In at least one of the bases, two measurements on both polarization states are required. Thus $4 \times 4= 16$ independent time resolved polarization sensitive two photon correlation measurements are needed in order to fully characterize the temporal evolution of the two photons polarization density matrix~\cite{james2001measurement}.
%The measured data are then used to construct the time dependent polarization density matrix
Considering only the polarization bases of Eq. \ref{EQ6_1}-\ref{EQ6_3} one gets 36 possible different two photon polarization sensitive probability rates, which divide essentially to 4 different cases (See Table \ref{TABLE1}).
In the 16 cases where $\theta_1=\theta_2=\pi/2$ oscillatory temporal dependence of the probability rate is expected with a relative phase which is given by $\phi_1+\phi_2$.  

%It is important to note here, that the exciton has a finite radiative lifetime. Thus EQ. \ref{EQ7} should describe the actual experimental time dependent correlation measurements after multiplication by a characteristic exponential decay model which takes into account the radiative decay of the exciton.  
%\begin{table*}[t]
\begin{table}[h]
%\centering
    \begin{tabular}{ c  c  c }
	\hline \hline
    Criteria                                                     & \# of cases & $p(t,P_1,P_2)/p_{X^0}(t)$            \\ \hline
    $\theta_1=\theta_2=0(\pi)$                 & 2          & $\frac{1}{2}$ \\ \hline
    $\theta_1=\pi (0)$, $\theta_2=0 (\pi)$ & 2          & 0                        \\ \hline
    $\theta_1=0~ or~ \pi$,$\theta_2=\pi/2$                                                           & 8           & $\frac{1}{4}$                        \\ \hline
    $\theta_1=\pi/2$,$\theta_2=0~ or~ \pi$                                                           & 8           & $\frac{1}{4}$                        \\ \hline
    $\theta_1=\theta_2=\pi/2$                    & 16          & $\frac{1}{4}\left[ 1+\cos\left(\phi_1+\phi_2+2\pi\frac{t}{T_P} \right)\right]$                         \\ \hline \hline
    \end{tabular}
\caption{Various two photon polarization sensitive emission probability rates.}
\label{TABLE1}
\end{table}

The sample was grown by molecular beam epitaxy on $[001]$-oriented GaAs substrate.
A layer of strain-induced InAs QDs was deposited in the center of an intrinsic GaAs layer. The GaAs layer was placed between two AlAs/GaAs distributed Bragg mirrors of quarter wavelength facilitating a microcavity. The microcavity design provides efficient collection of the light emitted due to recombination of QD confined electrons and holes from their respective lower energy levels. The QD is resonantly excited to the two photon biexciton absorption resonance using a synchronously pumped dye laser pulse with a repetition rate of $76\,\mathrm{MHz}$. The temporal width of the laser pulse is $\sim 12\,\mathrm{ps}$ and the spectral width is $\sim 100\,\mathrm{\mu eV}$. The resulting two photon emission is  measured using a polarization selective Hanbury Brown-Twiss (HBT) arrangement as schematically described in Fig.~\ref{fig:Fig2}(a). The emitted PL is divided by a non-polarizing beam splitter into two separated beams. A pair of liquid crystal variable retarders (LCVR) and a polarizing beam splitter on each beam is used to project the light on the desired polarization direction. The energy of the collected light on each beam is then filtered by a 1 meter long monochromator (MC) followed by a superconducting nanowire single photon detector (SNSPD) and a time-correlated single photon counter (TCSPC).

\section*{Results}
%In Fig.~\ref{fig:spectrum}(a) schematic description of the biexciton-exciton ($\XX_0\text{-}X_0$) cascade is shown together with the virtual biexciton level of the two photon excitation (TPE). The two co-linearly polarized photons paths ($H$ or $V$) are energetically distinguishable. 
In Fig.~\ref{fig:Fig2}(b) we present polarization sensitive PL spectra of the photoluminescence from the QD under resonant excitation into the biexciton two-photon absorption resonance~\cite{muller2014, brunner1994sharp}.
The energy of the exciting laser is tuned exactly between the $\XX^0$ and $X^0$ spectral line energies. Each of these spectral lines is composed of two cross linearly polarized components 
%with spectral width of roughly $50\,\mathrm{\murm eV}$ each and 
split  by  
$\Delta=34\,\mathrm{\mu eV}$ as can be clearly seen in Figs.~\ref{fig:Fig2}(c) and (d), which present the $X^0$ and $XX^0$ spectral lines, respectively,  on expanded energy scales. This energy splitting translates into an exciton precession time of $T_P=h/\Delta=122\,\mathrm{ps}$. 

Our setup was carefully designed to facilitate polarization sensitive time resolved correlation measurements between photons emitted from the $XX^0$ and the $X^0$ spectral lines. 
The correlation measurements were performed after deterministic generation of the $XX^0$ biexciton using a $\pi$-area pulse resonantly tuned into the
biexciton two-photon absorption resonance~\cite{brunner1994sharp,muller2014}. Prior to the generating pulse, a 4 ns long depletion pulse was used~\cite{Schmidgall2015}, verifying that the QD is empty from charges and dark excitons, ready for biexciton generation. The depletion pulse was produced using a continuous wave Ti:Sapphire laser light, temporally shaped by an electro optical modulator (EOM). The modulator was synchronized to the pulsed, synchronously pumped ps dye laser, which generated the biexciton (see Fig.~\ref{fig:Fig2}(a)).
 
\begin{figure}[t]
\includegraphics[width=\columnwidth]{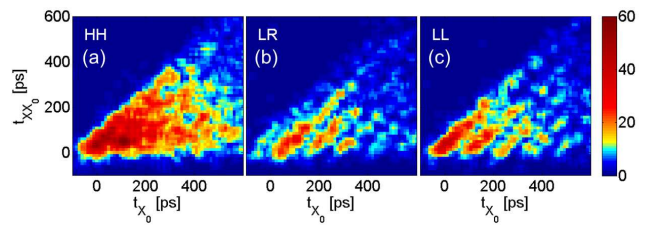}
\caption{\label{fig:pl_2d}The number of events (given by the color bar to the right) in which a biexciton and an exciton photons are detected as a function of their detection times as measured from the time of the pulsed excitation. In (a) the two photons are co-linearly polarized H, in (b) they are cross-circularly polarized (the biexciton L and the exciton R) and in (c) they are co-circularly polarized L.
%c) The number of events in which a biexciton and an exciton photon are simultaneously detected as a function %of the time difference between the detection of the first (biexciton) photon and the detection of the second (exciton) photon. Red [blue] curve presents the signal for cross- [co-] circularly 
%polarized photons, as obtained by integrating the signal in (a) [(b)]. 
%(d) The degree of circular polarization (as obtained by subtracting the signal of a) from that of b) and dividing by their sum) as a function of the two photon's detection times.  (e) The degree of the diagonal linear polarization (as obtained similarly to c) but from cross- and co-diagonally polarized two- photon time resolved correlation measurements) as a function of the two photon's detection times. (f) The degree of circular (red line) and diagonal (blue line) polarizations as a function of the time difference between the biexciton and exciton detections.  Note the quarter of a period phase difference between the two curves, and that the two degrees of polarizations do not decay during the radiative cascade. 
}
\end{figure}

In Fig.~\ref{fig:pl_2d}  the correlation measurements  are presented as two dimensional images showing by the false color scale the number of events in which two photons are detected as a function of their detection times after the excitation pulse. Fig.~\ref{fig:pl_2d} (a) shows the events when the two photons are H linearly co-polarized. Fig.~\ref{fig:pl_2d} (b) and (c), show the events when the two photons are cross and co-circularly polarized, respectively.
Fifteen more similar maps (not shown) were measured for different combinations of the two photon polarizations, providing the data required for constructing the temporal evolution of the two photon polarization density matrix. 
%In all these measurements we verified that the temporal evolution of the signal strongly depends on the time difference between the exciton detection time $t_{X^0}$ and the biexciton detection time $t_{XX^0}$. 
The temporal evolution of the signal integrated over the exciton detection time $t_{X^0}$ showed polarization independent simple exponential decay with a characteristic biexciton decay time of $260\pm10$ ps (not shown). 
Therefore, in order to increase the measurement statistics, we summed over the detection times of the biexciton $t_{XX^0}$ in Fig.~\ref{fig:pl_2d}  and obtained the number of coincidences as a function of the time difference between the photon detections.

\begin{figure}[t]
\includegraphics[width=\columnwidth]{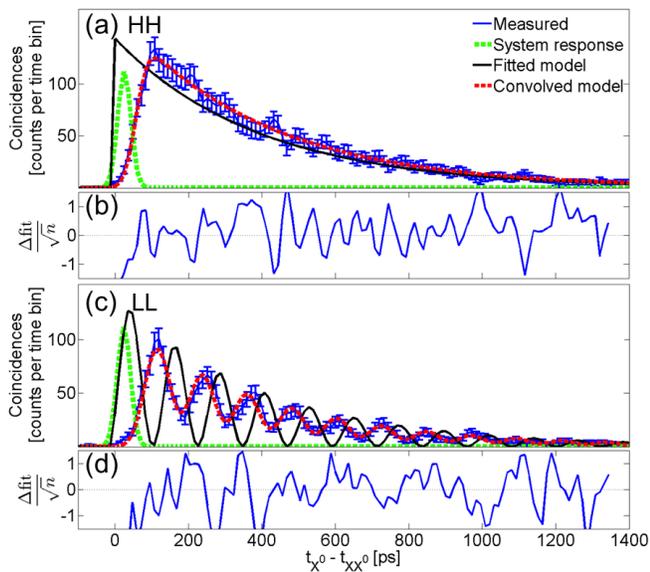}
\caption{\label{fig:fit} 
(a) The solid green line represents the measured temporal response function of the system. The blue marks with error bars represent the measured number of co-rectilinearly polarized two photon detection events as a function of the difference between their detection times. The black solid line represents the best fitted Eq. \ref{EQ7}  (which for this polarization projections is given by: $\frac{1}{2}p_{X^0}(t)$ ) and red solid line represents the model convolved by the system response function. 
(b) The difference between the convolved model and the measured data normalized by the experimental uncertainty. The quality of the fit is evident by the fact that the magnitude of the normalized difference does not exceed unity considerably. 
(c) and (d) are similar to (a) and (b), respectively, but for co-circularly polarized two photon events (note that in this case Eq. \ref{EQ7} is $\frac{1}{4}[1-\cos(\frac{2\pi t}{T_P})]p_{X^0}(t)$). 
}
\end{figure}

Typical measurements are presented by the blue symbols and error bars in Fig.~\ref{fig:fit}.
In Fig.~\ref{fig:fit} (a) the integrated data from Fig.~\ref{fig:pl_2d} (a) where the two photons are H co-linearly polarized are presented and in Fig.~\ref{fig:fit} (c) the data are obtained from Fig.~\ref{fig:pl_2d} (c) where the two photons are co-circularly polarized. In both figures, the green solid line presents the system temporal response as obtained by detecting the picosecond laser pulse, on both detectors. The system response is best-fitted by a Gaussian function with full width at half maximum of $42\,\mathrm{ps}$.

The solid black lines in Fig.~\ref{fig:fit} (a) and (c) represent the best fitted calculations using Eq. \ref{EQ7}. 
We note here that in fitting all the 16 different polarization sensitive correlation measurements, only one fitting parameter was used - $\tau_R$ the radiative lifetime of the exciton. We used $\tau_R=410\pm 10$ ps. The exciton precession period of $T_P=122\pm 12$ ps was directly obtained from the spectral measurement of the exciton fine structure.
Since all the correlation measurements were performed together, as the system automatically changed the polarization projections every minute, all the curves are automatically normalized and no attempt was made to fit their relative intensities. The overall coincidence accumulation rate of about $10^{-6}$ coincidences per pulse was also in agreement with the known light harvesting efficiency of our experimental setup~\cite{schmidgall2014deterministic}.    
The solid black lines represent the calculations using Eq. \ref{EQ7}.
The red lines overlaying the data are obtained by convolving the calculations by the system response function. 
Fig.~\ref{fig:fit} (b) and (d) show the difference between the data and the convolved calculations. These differences are within the experimental uncertainty and imply excellent agreement between the measured data and the convolved calculations.   

\begin{figure}[t]
\includegraphics[width=8.6cm]{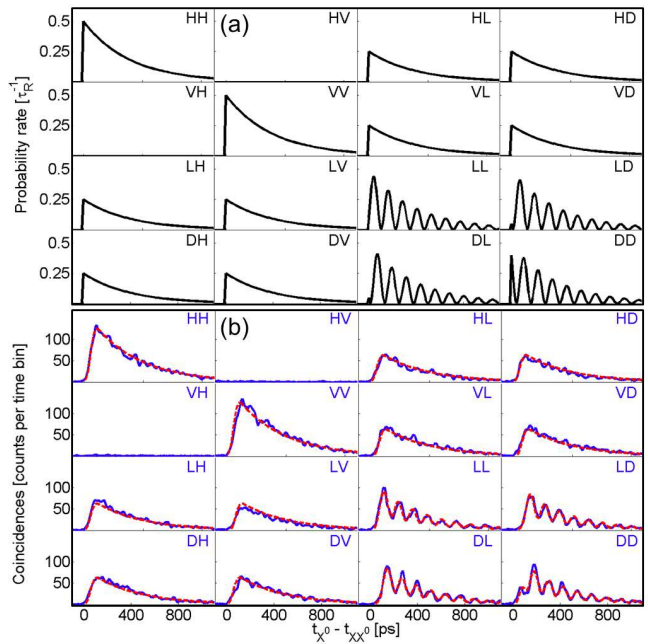}
\caption{\label{fig:data_fit_model} Sixteen different polarization sensitive two photon detection coincidences as a function of the time difference between the detection of the first and second photon. Solid black lines in (a) represent the calculated probability rates using Eq. \ref{EQ7}. Solid blue lines in (b) represent the measured coincidences rates,  and solid red lines in (b) represent the calculated probability rates convolved by the measured system temporal response as shown in Fig.~\ref{fig:fit}.  
The first (second) capital letter above each curve represents the polarization on which the biexciton (exciton) photon was projected. 
}
\end{figure}

In Fig.~\ref{fig:data_fit_model} (a) we present by solid black lines sixteen different polarization sensitive time resolved probability rates calculated using 
Eq.\ref{EQ7} (or Table \ref{TABLE1}).These sixteen different polarization projections are required for tomographic reconstruction of the two photon polarization density matrix. For these calculations we used $T_P=h/\Delta=122$ ps as deduced from the PL spectra, and $\tau_R= 410$ ps as the only fitted parameter. 
In Fig.~\ref{fig:data_fit_model} (b) solid blue lines represent the corresponding sixteen time resolved two photon coincidence rate measurements. Solid red lines in (b) overlaid on the measured data represent the calculations of Fig.~\ref{fig:data_fit_model} (a) convolved with the system response function from Fig.~\ref{fig:fit} (a) and (c). As can be seen in Fig.~\ref{fig:data_fit_model} (b), the agreement between the calculations and measurements is excellent. 

\begin{figure}[t]
\includegraphics[width=\columnwidth]{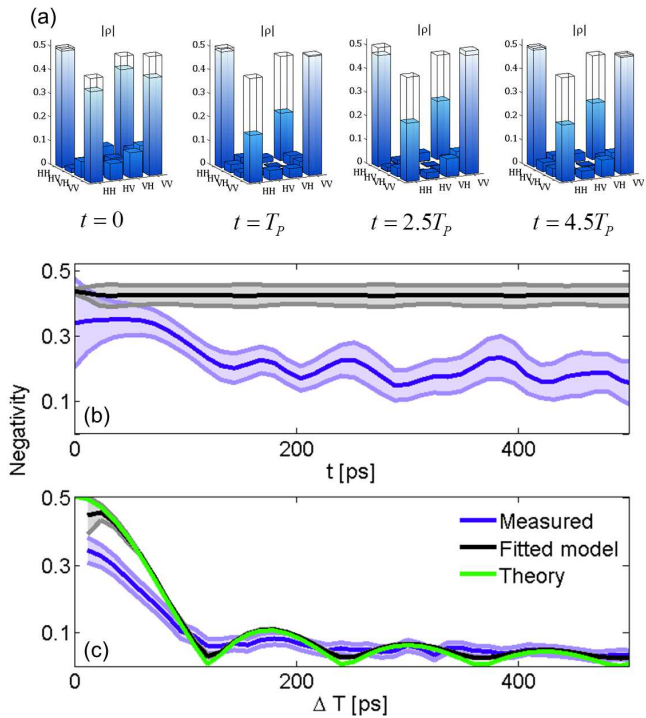}
\caption{\label{fig:density_matrix} 
(a) the absolute value of the two photon polarization density matrix for various times after the detection of the biexciton photon. Blue bars represent matrix elements obtained directly from the measured data, while empty bars represent matrix elements deduced from the fitted calculations, which consider the measured temporal response of the detectors. In both cases the temporal window width over which the data is integrated was set to $\Delta T =24$ ps.  
(b) The negativity of the two photon polarization density matrix as a function of the time difference between the exciton and biexciton detection times for temporal window width of $\Delta T =24$ ps. The blue line presents the negativity deduced directly from the measured raw data. The black line presents the negativity deduced from the fitted calculations considering the system response. %Shaded areas with corresponding colors represent uncertainties of one standard deviation.
(c) The negativity of the two photons polarization density matrix as a function of the temporal window width, $\Delta T$. The blue line presents the negativity deduced directly from the measured raw data. The black line presents the negativity deduced from the fitted calculations, considering the system response.
The theoretical dependence calculated by Eq. \ref{window} is represented by the green solid line.    
Shaded areas with corresponding colors in both (b) and (c) represent experimental uncertainties of one standard deviation.}
\end{figure}

The two-photon density matrix as function of time was reconstructed~\cite{james2001measurement} both from the measured data  and from the model calculations only.
The absolute values of four such matrices for various time differences after the detection of the biexciton photon are presented in Fig.~\ref{fig:density_matrix}(a). 
In both cases, the temporal window width, $\Delta T$, over which the data was integrated was set to $24$ ps.
In Fig.~\ref{fig:density_matrix}(a), blue bars  represent the measured (deduced from the data in Fig.~\ref{fig:data_fit_model} (b)) absolute values of the density matrix elements. Empty bars represent the matrix elements obtained from the fitted model calculations (deduced from Fig.~\ref{fig:data_fit_model} (b)), which consider the measured temporal response of our detectors. 

The negativity of the density matrix as a function of time between the photons is presented in Fig.~\ref{fig:density_matrix}(b), in blue for the as measured matrices and in black for the calculated ones, considering the detectors temporal response. 
Here as well, the temporal window width $\Delta T$, was set to $24$ ps.

Fig.~\ref{fig:density_matrix}(c) presents the negativity of the two photon polarization density matrix as a function of the temporal window width, $\Delta T$. The solid blue line in the figure presents the negativity as deduced directly from the measured raw data. The solid black line presents the negativity deduced from the fitted calculations, considering the detectors temporal response. Shaded areas with corresponding colors in both Fig.~\ref{fig:density_matrix}(b) and (c) represent uncertainties of one standard deviation. Note the excellent agreement between the solid black line and the theoretical dependence predicted by Eq. \ref{window}, represented in Fig.~\ref{fig:density_matrix}(c) by the green solid line.    

\section*{Summary}
In summary, we have demonstrated experimentally that the biexciton-exciton radiative cascade in semiconductor quantum dots is an excellent deterministic source of maximally polarization entangled photon pairs. We showed that: a) The polarization states of the emitted two photons are maximally entangled during the whole radiative decay. b) The measured degree of entanglement between the polarization states of the two photons depends only on the temporal resolution by which the time difference between the two photon emissions is determined. 
%c) If this resolution is considerably better than the exciton precession time (inverse of its fine structure splitting) the measured degree of entanglement will be close to maximal. 
A relatively simple arrangement, which provides periodic retardation change compatible with the exciton precession period, can therefore transform any quantum dot, into an on demand source of maximally entangled photons \cite{specht2009phase, jones2006photon, wang2010demand}. 

\begin{acknowledgments}
We thank N. H. Lindner, O. Kenneth and J. A. Avron for helpful discussions and P.M. Petroff for the sample. The support of the Israeli Science Foundation (ISF), the Technion's RBNI and the Israeli Focal Technology Area on ``Nanophotonics for Detection'' is gratefully acknowledged. This project has also received funding from the European Research Council (ERC) under the European
Union’s Horizon 2020 research and innovation programme (grant agreement No 695188).
\end{acknowledgments}

\bibliographystyle{unsrt}
\bibliography{master_bibtex_db}

\begin{thebibliography}{10}

\bibitem{benson2000}
O.~Benson, C.~Santori, M.~Pelton, and Y.~Yamamoto.
\newblock Regulated and entangled photons from a single quantum dot.
\newblock {\em Phys. Rev. Lett.}, 84:2513--2516, Mar 2000.

\bibitem{akopian2006}
N.~Akopian, N.~H. Lindner, E.~Poem, Y.~Berlatzky, J.~Avron, D.~Gershoni, B.~D.
  Gerardot, and P.~M. Petroff.
\newblock Entangled photon pairs from semiconductor quantum dots.
\newblock {\em Phys. Rev. Lett.}, 96:130501, Apr 2006.

\bibitem{sellinart2010}
A.~Dousse, J.~Suffczynski, A.~Beveratos, O.~Krebs, A.~Lemaitre, I.~Sagnes,
  J.~Bloch, P.~Voisin, and P.~Senellart.
\newblock Ultrabright source of entangled photon pairs.
\newblock {\em Nature}, 466:217--220, 2010.

\bibitem{bounouar2015phonon}
S.~Bounouar, M.~M{\"u}ller, A.~M. Barth, M.~Gl{\"a}ssl, V.~M. Axt, and
  P.~Michler.
\newblock Phonon-assisted robust and deterministic two-photon biexciton
  preparation in a quantum dot.
\newblock {\em Physical Review B}, 91(16):161302, 2015.

\bibitem{muller2014}
M.~M{\"u}ller, S.~Bounouar, K.~D. J{\"o}ns, M.~Gl{\"a}ssl, and P.~Michler.
\newblock On-demand generation of indistinguishable polarization-entangled
  photon pairs.
\newblock {\em Nature Photon.}, 8:224, 2014.

\bibitem{stevenson2008evolution}
R.~M. Stevenson, A.~J. Hudson, A.~J. Bennett, R.~J. Young, C.~A. Nicoll, D.~A.
  Ritchie, and A.~J. Shields.
\newblock Evolution of entanglement between distinguishable light states.
\newblock {\em Physical Review Letters}, 101(17):170501, 2008.

\bibitem{young2005inversion}
R.~J. Young, R.~M. Stevenson, A.~J. Shields, P.~Atkinson, K.~Cooper, D.~A.
  Ritchie, K.~M. Groom, A.~I. Tartakovskii, and M.~S. Skolnick.
\newblock Inversion of exciton level splitting in quantum dots.
\newblock {\em Physical Review B}, 72(11):113305, 2005.

\bibitem{versteegh2014observation}
M.~A.~M. Versteegh, M.~E. Reimer, K.~D. J{\"o}ns, D.~Dalacu, P.~J. Poole,
  A.~Gulinatti, A.~Giudice, and V.~Zwiller.
\newblock Observation of strongly entangled photon pairs from a nanowire
  quantum dot.
\newblock {\em Nature communications}, 5, 2014.

\bibitem{zhang2015high}
J.~Zhang, J.~S. Wildmann, F.~Ding, R.~Trotta, Y.~Huo, E.~Zallo, D.~Huber,
  A.~Rastelli, and O.~G. Schmidt.
\newblock High yield and ultrafast sources of electrically triggered
  entangled-photon pairs based on strain-tunable quantum dots.
\newblock {\em Nature communications}, 6, 2015.

\bibitem{bennett2010electric}
A.~J. Bennett, M.~A. Pooley, R.~M. Stevenson, M.~B. Ward, R.~B. Patel, A.~Boyer
  de~La~Giroday, N.~Sk{\"o}ld, I.~Farrer, C.~A. Nicoll, D.~A. Ritchie, et~al.
\newblock Electric-field-induced coherent coupling of the exciton states in a
  single quantum dot.
\newblock {\em Nature Physics}, 6(12):947--950, 2010.

\bibitem{young2006improved}
R.~J. Young, R.~M. Stevenson, P.~Atkinson, K.~Cooper, D.~A. Ritchie, and A.~J.
  Shields.
\newblock Improved fidelity of triggered entangled photons from single quantum
  dots.
\newblock {\em New Journal of Physics}, 8(2):29, 2006.

\bibitem{stevenson2006magnetic}
R.~M. Stevenson, R.~J. Young, P.~See, D.~G. Gevaux, K.~Cooper, P.~Atkinson,
  I.~Farrer, D.~A. Ritchie, and A.~J. Shields.
\newblock Magnetic-field-induced reduction of the exciton polarization
  splitting in inas quantum dots.
\newblock {\em Physical Review B}, 73(3):033306, 2006.

\bibitem{Schmidgall2015}
E.~R. Schmidgall, I.~Schwartz, D.~Cogan, L.~Gantz, T.~Heindel, S.~Reitzenstein,
  and D.~Gershoni.
\newblock All-optical depletion of dark excitons from a semiconductor quantum
  dot.
\newblock {\em Applied Physics Letters}, 106(19):193101, 2015.

\bibitem{jones2006photon}
N.~S. Jones and T.~M. Stace.
\newblock Photon frequency-mode matching using acousto-optic frequency beam
  splitters.
\newblock {\em Physical Review A}, 73(3):033813, 2006.

\bibitem{wang2010demand}
X.~B. Wang, C.~X. Yang, and Y.~B. Liu.
\newblock On-demand entanglement source with polarization-dependent frequency
  shift.
\newblock {\em Applied Physics Letters}, 96(20):201103, 2010.

\bibitem{brunner1994sharp}
K.~Brunner, G.~Abstreiter, G.~B{\"o}hm, G.~Tr{\"a}nkle, and G.~Weimann.
\newblock Sharp-line photoluminescence and two-photon absorption of
  zero-dimensional biexcitons in a gaas/algaas structure.
\newblock {\em Physical Review Letters}, 73(8):1138, 1994.

\bibitem{stufler2006two}
S.~Stufler, P.~Machnikowski, P.~Ester, M.~Bichler, V.~M. Axt, T.~Kuhn, and
  A.~Zrenner.
\newblock Two-photon rabi oscillations in a single in x ga 1- x as/ ga as
  quantum dot.
\newblock {\em Physical Review B}, 73(12):125304, 2006.

\bibitem{jayakumar2013deterministic}
H.~Jayakumar, A.~Predojevi{\'c}, T.~Huber, T.~Kauten, G.~S. Solomon, and
  G.~Weihs.
\newblock Deterministic photon pairs and coherent optical control of a single
  quantum dot.
\newblock {\em Physical review letters}, 110(13):135505, 2013.

\bibitem{gammon1996}
D.~Gammon, E.~S. Snow, B.~V. Shanabrook, D.~S. Katzer, and D.~Park.
\newblock Fine structure splitting in the optical spectra of single gaas
  quantum dots.
\newblock {\em Phys. Rev. Lett.}, 76:3005--3008, Apr 1996.

\bibitem{kulakovskii1999fine}
V.~D. Kulakovskii, G.~Bacher, R.~Weigand, T.~K{\"u}mmell, A.~Forchel,
  E.~Borovitskaya, K.~Leonardi, and D.~Hommel.
\newblock Fine structure of biexciton emission in symmetric and asymmetric
  cdse/znse single quantum dots.
\newblock {\em Physical Review Letters}, 82(8):1780, 1999.

\bibitem{benny2011prl}
Y.~Benny, S.~Khatsevich, Y.~Kodriano, E.~Poem, R.~Presman, D.~Galushko, P.~M.
  Petroff, and D.~Gershoni.
\newblock Coherent optical writing and reading of the exciton spin state in
  single quantum dots.
\newblock {\em Phys. Rev. Lett.}, 106:040504, Jan 2011.

\bibitem{poem2011}
E.~Poem, O.~Kenneth, Y.~Kodriano, Y.~Benny, S.~Khatsevich, J.~E. Avron, and
  D.~Gershoni.
\newblock Optically induced rotation of an exciton spin in a semiconductor
  quantum dot.
\newblock {\em Phys. Rev. Lett.}, 107:087401, Aug 2011.

\bibitem{ward2014coherent}
M.~B. Ward, M.~C. Dean, R.~M. Stevenson, A.~J. Bennett, D.~JP Ellis, K.~Cooper,
  I.~Farrer, C.~A. Nicoll, D.~A. Ritchie, and A.~J. Shields.
\newblock Coherent dynamics of a telecom-wavelength entangled photon source.
\newblock {\em Nature communications}, 5, 2014.

\bibitem{peres1996separability}
A.~Peres.
\newblock Separability criterion for density matrices.
\newblock {\em Physical Review Letters}, 77(8):1413, 1996.

\bibitem{scully1997quantum}
M.~O. Scully and M.~S. Zubairy.
\newblock {\em Quantum optics}.
\newblock Cambridge university press, 1997.

\bibitem{schmidgall2014deterministic}
E.~R. Schmidgall, I.~Schwartz, L.~Gantz, D.~Cogan, S.~Raindel, and D.~Gershoni.
\newblock Deterministic generation of a quantum-dot-confined triexciton and its
  radiative decay via three-photon cascade.
\newblock {\em Physical Review B}, 90(24):241411, 2014.

\bibitem{james2001measurement}
D.~James, P.~G. Kwiat, W.~J. Munro, and A.~G. White.
\newblock Measurement of qubits.
\newblock {\em Physical Review A}, 64(5):052312, 2001.

\bibitem{specht2009phase}
H.~P. Specht, J.~Bochmann, M.~M{\"u}cke, Be. Weber, E.~Figueroa, D.~L.
  Moehring, and G.~Rempe.
\newblock Phase shaping of single-photon wave packets.
\newblock {\em Nature Photonics}, 3(8):469--472, 2009.

\end{thebibliography}

\end{document}